\documentclass[prl,twocolumn,superscriptaddress,showpacs,a4paper]{revtex4}
\usepackage{epsfig}
\usepackage{amsmath}
\usepackage{graphicx}
\usepackage{bm}

\begin{document}

\title{Optomechanical entanglement between a movable mirror and a cavity field}

\author{D. Vitali}\affiliation{Dipartimento di Fisica, Universit\`a di Camerino, I-62032 Camerino (MC), Italy}
\author{S. Gigan}\affiliation{Institut f\"{u}r Experimentalphysik, Universit\"{a}t
Wien, Boltzmanngasse 5, 1090 Wien, Austria}
\author{A. Ferreira}\affiliation{Institut f\"{u}r Experimentalphysik, Universit\"{a}t
Wien, Boltzmanngasse 5, 1090 Wien, Austria}\affiliation{Faculdade
Ci\^{e}ncias Universidade do Porto,  Rua do Campo Alegre 687
4169-007, Porto, Portugal}
\author{H. R. B\"ohm}\affiliation{Institut f\"{u}r Experimentalphysik, Universit\"{a}t
Wien, Boltzmanngasse 5, 1090 Wien, Austria}
\author{P. Tombesi}\affiliation{Dipartimento di Fisica, Universit\`a di Camerino, I-62032 Camerino (MC), Italy}
\author{A. Guerreiro}\affiliation{Faculdade Ci\^{e}ncias Universidade do
Porto,  Rua do Campo Alegre 687 4169-007, Porto,
Portugal}\author{V. Vedral}\affiliation{Institut f\"{u}r
Experimentalphysik, Universit\"{a}t Wien, Boltzmanngasse 5, 1090
Wien, Austria}\affiliation{The School of Physics and Astronomy,
University of Leeds, Leeds, LS2 9JT, UK}
\author{A. Zeilinger}\affiliation{Institut f\"{u}r Experimentalphysik, Universit\"{a}t
Wien, Boltzmanngasse 5, 1090 Wien, Austria}\affiliation{Institute
for Quantum Optics and Quantum Information (IQOQI), Austrian
Academy of Sciences, Boltzmanngasse 3, 1090 Wien, Austria}
\author{M. Aspelmeyer}\affiliation{Institut f\"{u}r Experimentalphysik, Universit\"{a}t
Wien, Boltzmanngasse 5, 1090 Wien, Austria}\affiliation{Institute
for Quantum Optics and Quantum Information (IQOQI), Austrian
Academy of Sciences, Boltzmanngasse 3, 1090 Wien, Austria}

\date{\today}

\begin{abstract}
We show how stationary entanglement between an optical
cavity field mode and a macroscopic vibrating mirror can be
generated by means of radiation pressure. We also show how the
generated optomechanical entanglement can be quantified and we
suggest an experimental readout-scheme to fully characterize the
entangled state. Surprisingly, such optomechanical entanglement
is shown to persist for environment temperatures above 20K using
state-of-the-art experimental parameters.
\end{abstract}

\pacs{03.67.Mn, 42.50.Lc, 05.40.Jc}

\maketitle

Entanglement, ``\textit{the} characteristic trait of quantum
mechanics''~\cite{Schrodinger35}, has raised widespread interest
in different branches of physics. It provides insight into the
fundamental structure of physical reality~\cite{Bell64} and it
has become a basic resource for many quantum information
processing schemes~\cite{entmicro1}. So far entanglement has been
experimentally prepared and manipulated using microscopic quantum
systems such as photons, atoms and
ions~\cite{entmicro1,entmicro2}. Nothing in the principles of
quantum mechanics prevents macroscopic systems to attain
entanglement. However, the answer to the question as to what
extent entanglement should hold when going towards ``classical''
systems is yet unknown~\cite{Leggett}. Therefore it is of crucial
importance to investigate the possibilities to obtain entangled
states of macroscopic systems~\cite{Armour02} and to study the
robustness of entanglement against temperature~\cite{Vedral}.
Experiments in this direction include single-particle
interference of macro-molecules~\cite{Hackermueller}, the
demonstration of entanglement between collective spins of atomic
ensembles~\cite{Polzik}, and of entanglement in
Josephson-junction qubits~\cite{Berkley}. Mechanical oscillators
are of particular interest since they resemble a prototype of
``classical'' systems. Thanks to the fast-developing field of
microfabrication, micro- or nano-mechanical oscillators can now be
prepared and controlled to a very high precision \cite{Roukes03}.
In addition, several theoretical proposals exist that suggest how
to reach the quantum regime for such systems \cite{Pinard0}.
Experimentally, quantum limited measurements have been developed
that could allow ground state detection~\cite{LaHaye}. However,
quantum effects in mechanical oscillators have not been
demonstrated to date.

Optomechanical coupling via radiation pressure~\cite{Braginsky} is a
promising approach to prepare and manipulate quantum states of
mechanical oscillators. Proposals range from the quantum state
transfer from light to a mechanical oscillator to entangling two
such oscillators \cite{Braunstein,Marshall03,pinard2,Tombesi}. In
this paper we propose an experimental scheme to create and probe
optomechanical entanglement between a light field and a mechanical
oscillator. This is achieved using a bright laser field that
resonates inside a cavity and couples to the position and momentum
of a moving (micro)mirror. The proposal is based on feasible
experimental parameters in accordance with current state of the art
optics and microfabrication. In contrast to other
proposals~\cite{Braunstein,pinard2} it neither requires
non-classical states of light nor temperatures close to the
oscillator's ground state. Entanglement is shown to persist above a
temperature of 20K. We begin by modelling the system and its
coupling to the environment by using the standard Langevin
formalism. Then we solve the dynamics and quantify the entanglement
generated in the stationary state. Finally we discuss a suitable
experimental apparatus capable of measuring the entanglement.

We consider an optical Fabry-Perot cavity in which one of the
mirrors is much lighter than the other, so that it can move under
the effect of the radiation pressure force. The motion of the
mirror is described by the excitation of several degrees of
freedom which have different resonant frequencies. However, a
single frequency mode can be considered when a bandpass filter in
the detection scheme is used \cite{Pinard} and mode-mode coupling is negligible. Therefore we
will consider a single mechanical mode of the mirror only,
which can be modeled as an harmonic oscillator with
frequency $w_m$. The Hamiltonian of the system reads \cite{law}
\begin{eqnarray} \label{ham0}
&&\mathcal{H}=\hbar w_c a^{\dag}a + \frac{\hbar w_m}{2}(p^2+q^2)
-\hbar G_0 a^{\dag}a q \\
&& \nonumber + \imath\hbar E (e^{-\imath w_{0} t}a^{\dag}-e^{\imath
w_{0} t}a),
\end{eqnarray}
where $q$ and $p$ ($[q,p]=i$) are the dimensionless position and
momentum operators of the mirror, $a$ and $a^{\dag}$
($[a,a^{\dag}]=1$) are the annihilation and creation operators of
the cavity mode with frequency $w_c$ and decay rate $\kappa$, and
$G_0=(w_c/L)\sqrt{\hbar/m w_m}$ is the coupling coefficient, with
$L$ the cavity length in the absence of the intracavity field and
$m$ the effective mass of the mechanical mode \cite{Pinard}. The
last two terms in Eq.~(\ref{ham0}) describe the driving laser
with frequency $w_0$ and $E$ is related to the input laser power
$P$ by $|E|=\sqrt{2P \kappa/\hbar w_0}$.

A proper analysis of the system must include photon losses in the
cavity and the Brownian noise acting on the mirror. This can be
accomplished by considering the following set of nonlinear Langevin equations,
written in the interaction picture with respect to $\hbar w_0 a^{\dag}a$
\begin{subequations}
\label{nonlinlang}
\begin{eqnarray}
 \dot{q}&=&w_m p, \\  \dot{p}&=&-w_m q - \gamma_m
p + G_0  a^{\dag}a +
\xi, \\  \dot{a}&=&-(\kappa+i\Delta_0)a +i G_0 a q +E+\sqrt{2\kappa} a^{in},
\end{eqnarray}
\end{subequations}
where $\Delta_0=w_c-w_0$ and $\gamma_m$ is the mechanical damping rate. We have introduced the vacuum radiation input noise
$a^{in}$, whose only nonzero correlation function is \cite{milwal}
\begin{equation}\label{input}
  \langle a^{in}(t)a^{in,\dag}(t')\rangle =\delta (t-t'),
\end{equation}
and the Hermitian Brownian noise operator $\xi$, with correlation function \cite{Giovannetti01}
\begin{equation}\label{browncorre}
\left \langle \xi(t) \xi(t')\right \rangle = \frac{\gamma_m}{w_m} \int
  \frac{d\omega}{2\pi} e^{-i\omega(t-t')} \omega \left[\coth\left(\frac{\hbar \omega}{2k_BT}\right)+1\right]
\end{equation}
($k_B$ is the Boltzmann constant and $T$ is the mirror temperature).
We can always rewrite each Heisenberg operator as a c-number steady
state value plus an additional fluctuation operator with zero mean
value, $a = \alpha_s + \delta a $, $ q = q_s +  \delta q $, $p = p_s
+ \delta p$. Inserting these expressions into the Langevin equations
of Eqs.~(\ref{nonlinlang}), these latter decouple into a set of
nonlinear algebraic equations for the steady state values and a set
of quantum Langevin equations for the fluctuation
operators~\cite{fabre}. The steady state values are given by $ p_s
=0$, $q_s = G_0 |\alpha_s|^2/w_m$, $\alpha_s = E/(\kappa+i \Delta)$,
where the latter equation is in fact a nonlinear equation
determining the stationary intracavity field amplitude $\alpha_s$,
since the effective cavity detuning $\Delta$, including radiation
pressure effects, is given by $\Delta = \Delta_0- G_0^2
|\alpha_s|^2/w_m$. The parameter regime relevant for generating
optomechanical entanglement is that with a very large input power
$P$, i.e., when $|\alpha_s| \gg 1$. In this case, one can safely
neglect the nonlinear terms $\delta a^{\dag} \delta a$ and $ \delta
a \delta q$ and gets the linearized Langevin equations
\begin{subequations}
\label{lle}
\begin{eqnarray}
\delta \dot{q}&=&w_m \delta p, \\  \delta \dot{p}&=&-w_m \delta q - \gamma_m
\delta p + G \delta X +\xi, \\
 \delta \dot{X}&=&-\kappa \delta X+\Delta \delta Y +\sqrt{2\kappa} X^{in},  \\
 \delta \dot{Y}&=&-\kappa \delta Y-\Delta \delta X +G\delta q +\sqrt{2\kappa} Y^{in},
\end{eqnarray}
\end{subequations}
where we have chosen the phase reference of the cavity field so that
$\alpha_s$ is real, we have defined the cavity field quadratures
$\delta X\equiv(\delta a+\delta a^{\dag})/\sqrt{2}$ and $\delta
Y\equiv(\delta a-\delta a^{\dag})/i\sqrt{2}$, and the corresponding
Hermitian input noise operators
$X^{in}\equiv(a^{in}+a^{in,\dag})/\sqrt{2}$ and
$Y^{in}\equiv(a^{in}-a^{in,\dag})/i\sqrt{2}$. What is relevant is
that the quantum fluctuations of the field and the oscillator are
now coupled by the much larger \emph{effective} optomechanical
coupling $G \equiv G_0 \alpha_s \sqrt{2}$, so that the generation of
significant optomechanical entanglement becomes possible.

When the system is stable it reaches a unique steady state,
independently of the initial condition. Since the quantum noises
$\xi$ and $a^{in}$ are zero-mean quantum Gaussian noises and the
dynamics is linearized, the quantum steady state for the
fluctuations is a zero-mean bipartite Gaussian state, fully
characterized by its $4 \times 4 $ correlation matrix $
V_{ij}=\left(\langle u_i(\infty)u_j(\infty)+
u_j(\infty)u_i(\infty)\rangle\right)/2$, where $u^{T}(\infty)
=(\delta q(\infty), \delta p(\infty),\delta X(\infty), \delta
Y(\infty))$ is the vector of continuous variables (CV) fluctuation
operators at the steady state ($t \to \infty$). Defining the vector
of noises $n^{T}(t) =(0, \xi(t),\sqrt{2\kappa}X^{in}(t),
\sqrt{2\kappa}Y^{in}(t))$ and the matrix
\begin{equation}\label{dynmat}
  A=\left(\begin{array}{cccc}
    0 & w_m & 0 & 0 \\
     -w_m & -\gamma_m & G & 0 \\
    0 & 0 & -\kappa & \Delta \\
    G & 0 & -\Delta & -\kappa
  \end{array}\right),
\end{equation}
Eqs.~(\ref{lle}) can be written in compact form as $\dot{u}(t)=A u(t)+n(t)$,
whose solution is
$ u(t)=M(t) u(0)+\int_0^t ds M(s) n(t-s)$,
where $M(t)=\exp\{A t\}$. The system is stable and reaches its steady state when all the eigenvalues
of $A$ have negative real parts so that $M(\infty)=0$.
The stability conditions can be derived by applying the Routh-Hurwitz criterion \cite{grad}, yielding the following two nontrivial
conditions on the system parameters,
\begin{subequations}
\label{stab}
\begin{eqnarray}
&& 2 \gamma_m \kappa \left[\Delta^4 +\Delta^2(\gamma_m^2+2\gamma_m \kappa +2 \kappa^2-2w_m^2)\right. \\
&&\left.+\left(\gamma_m \kappa +\kappa^2+w_{m}^{2}\right)^2\right]+ w_m G^2 \Delta
(\gamma_m+2\kappa)^2 > 0 , \nonumber \\
&& w_m^2\left(\Delta^2+\kappa^2\right)-w_m G^2 \Delta > 0,
\end{eqnarray}
\end{subequations}
which will be considered to be satisfied from now on. When the
system is stable one gets
\begin{equation} \label{cm2}
V_{ij}=\sum_{k,l}\int_0^{\infty} ds \int_0^{\infty}ds' M_{ik}(s) M_{jl}(s')\Phi_{kl}(s-s'),
\end{equation}
where
$\Phi_{kl}(s-s')=\left(\langle n_k(s)n_l(s')+ n_l(s')n_k(s)\rangle\right)/2$
is the matrix of the stationary noise correlation functions. Due to Eq.~(\ref{browncorre}), the mirror Brownian noise $\xi(t)$
is not delta-correlated
and therefore does not describe a Markovian process~\cite{Giovannetti01}. However, quantum effects
are achievable only using oscillators with a large mechanical quality factor $\mathcal{Q}=w_m /\gamma_m \gg 1$. In this limit,
$\xi(t)$ becomes delta-correlated~\cite{benguria},
\begin{equation}\label{browncorre6}
\left \langle \xi(t) \xi(t')+\xi(t') \xi(t)\right \rangle/2 \simeq \gamma_m
\left(2\bar{n}+1\right) \delta(t-t'),
\end{equation}
where $\bar{n}=\left(\exp\{\hbar w_m/k_BT\}-1\right)^{-1}$ is the mean thermal excitation number, and one recovers a Markovian process.
As a consequence, and using the fact that the three components of $n(t)$ are uncorrelated, we get
$\Phi_{kl}(s-s')= D_{kl} \delta(s-s')$,
where $D = \mathrm{Diag }
[0,\gamma_m (2 \bar{n}+1),\kappa,\kappa]$ is a diagonal matrix, and Eq.~(\ref{cm2}) becomes
$ V =\int_0^{\infty} ds  M(s)D M(s)^{T}$.
When the stability conditions are satisfied, $M(\infty)=0$ and
one gets the following equation for the
steady-state CM,
\begin{equation} \label{lyap}
AV+VA^{T}=-D.
\end{equation}
Eq.~(\ref{lyap}) is a linear equation for $V$ and can be
straightforwardly solved, but the general exact expression is too
cumbersome and will not be reported here. In order to establish the
conditions under which the optical mode and the mirror vibrational
mode are entangled we consider the logarithmic negativity
$E_{\mathcal{N}}$, a quantity which has been already proposed as a
measure of entanglement \cite{werner}. In the CV case
$E_{\mathcal{N}}$ can be defined as \cite{Salerno1}
\begin{equation}
E_{\mathcal{N}}=\max [0,-\ln 2\eta ^{-}],  \label{logneg}
\end{equation}
where
$
\eta ^{-}\equiv 2^{-1/2}\left[ \Sigma (V)- \left[ \Sigma (V)^{2}-4\det
V\right] ^{1/2}\right] ^{1/2}$,
with $\Sigma (V)\equiv \det A+\det B-2\det C$, and we have used the $2\times2$ block form
of the CM
\begin{equation}
V\equiv \left(
\begin{array}{cc}
A & C \\
C^{T} & B
\end{array}
\right) .  \label{blocks}
\end{equation}
Therefore, a Gaussian state is entangled if and only if $ \eta
^{-}<1/2$, which is equivalent to Simon's necessary and sufficient
entanglement non-positive partial transpose criterion for Gaussian
states \cite{simon}, which can be written as $4\det V < \Sigma
-1/4$.

We have made a careful analysis in a wide parameter range and found
a parameter region very close to that of recently performed
optomechanical experiments \cite{gigan06}, for which a significative
amount of entanglement is achievable. Fig.~1 shows $E_{\mathcal{N}}$
versus the normalized detuning $\Delta/w_m$ for two different
masses, $5$ and $50$ ng: optomechanical entanglement is present only
within a finite interval of values of $\Delta$ around $\Delta \simeq
w_m$. The robustness of such an entanglement with respect to the
mirror's environmental temperature is shown in Fig.~2. The relevant
result is that for the $5$ ng mirror optomechanical entanglement
persists for temperatures above 20K, which is several orders of
magnitude larger than the ground state temperature of the mechanical
oscillator. For the $50$ ng mirror entanglement vanishes at lower
temperatures (Fig.~2). Figs.~1-2 refer to ${\cal Q}= 10^5$, but we
found that entanglement persists even for ${\cal Q}\simeq 10^4$,
although it becomes much less robust against temperature. In this
case, entanglement persists up to $3(1)$ K for a $5 (50)$ ng mirror.

\begin{figure}[htb]
\centerline{\includegraphics[width=0.39\textwidth]{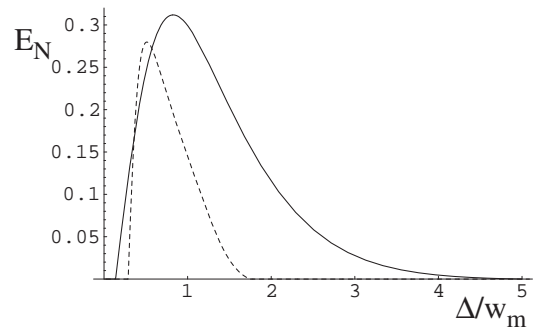}}
\caption{Plot of the logarithmic negativity $E_{\mathcal{N}}$ as a
function of the normalized detuning $\Delta/w_m$ in the case of an
optical cavity of length $L=1$ mm, driven by a laser with wavelength
$810$ nm and power $P=50$ mW. The mechanical oscillator has a
frequency $w_m/2\pi=10$ MHz, a damping rate $\gamma_m/2\pi=100$ Hz,
and its temperature is $T=400$ mK. Full line refers to a mass $m=5$
ng, and finesse $\mathcal{F}=1.07 \times 10^4$, while the dashed
line refers to a mass $m=50$ ng and finesse $\mathcal{F}=3.4 \times
10^4$.} \label{fig2}
\end{figure}

\begin{figure}[htb]
\centerline{\includegraphics[width=0.39\textwidth]{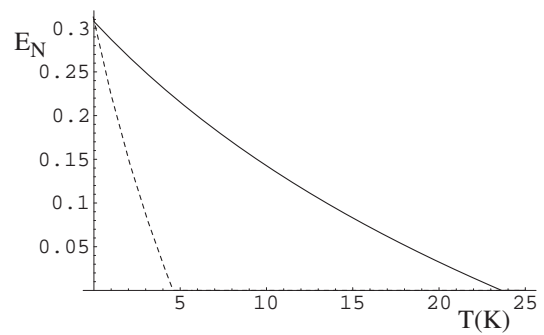}}
\caption{Plot of the logarithmic negativity $E_{\mathcal{N}}$ versus
the mirror temperature. The full line refers to a mass $m=5$ ng,
detuning $\Delta=w_m$ and finesse $\mathcal{F}=1.07 \times 10^4$;
the dashed line refers to a mass $m=50$ ng, $\Delta=w_m/2$, and
finesse $\mathcal{F}=3.4 \times 10^4$. The other parameters are
those of Fig.~1.} \label{fig4}
\end{figure}

We finally discuss the experimental detection of the generated
optomechanical entanglement. In order to measure $E_{\mathcal{N}}$
at the steady state, one has to measure all the ten independent
entries of the correlation matrix V. This has been recently
experimentally realized~\cite{laurat} for the case of two entangled
optical modes at the output of a parametric oscillator. In our case,
the measurement of the field quadratures of the cavity mode can be
straightforwardly performed by homodyning the cavity output using a
local oscillator with an appropriate phase. Measuring the mechanical
mode is less straightforward. However, if we consider a second
Fabry-Perot cavity $C_2$, adjacent to the first one and formed by
the movable mirror and a third fixed mirror (see Fig.~3), it is
possible to adjust the parameters of $C_2$ so that both position and
momentum of the mirror can be measured by homodyning the $C_2$
output. In fact, assuming that the movable mirror has unit
reflectivity at both sides so that there is no light coupling the
two cavities, the annihilation operator of the second cavity, $a_2$,
obeys an equation analogous to the linearized version of
Eq.~(\ref{nonlinlang}c),
\begin{equation}
\label{mode2} \delta \dot{a}_2 =-(\kappa_2+i\Delta_2)\delta a_2 +i G_2 \alpha_2 \delta q +\sqrt{2\kappa_2} a_{2}^{in}(t),
\end{equation}
where $\kappa_2$, $\Delta_2$, $\alpha_2$, $a_{2}^{in}(t)$ are the
bandwidth, the effective detuning, the intracavity field amplitude,
and the input noise of $C_2$, respectively. Moreover,
$G_2=(w_{c2}/L_2)\sqrt{\hbar/mw_m}$, where $w_{c2}$ and $L_2$ are
the frequency and the length of $C_2$. The presence of the second
cavity affects the mirror dynamics, which is no more exactly
described by Eqs.~(\ref{lle}). However, if $C_2$ is driven by a much
weaker intracavity field so that $|\alpha_2| \ll |\alpha_s|$, its
back-action on the mechanical mode can be neglected and the relevant
dynamics is still well described by Eqs.~(\ref{lle}).

\begin{figure}[htb]
\centerline{\includegraphics[width=0.45\textwidth]{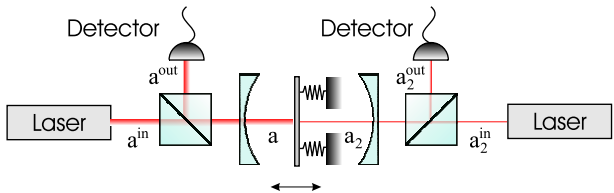}}
\caption{Schematic description of the proposed experiment, including
the second Fabry-Perot cavity on the right for the detection of the
mechanical motion.} \label{fig3}
\end{figure}

If we now choose parameters so that $\Delta_2=w_m \gg k_2, G_2 |\alpha_2|$, we can rewrite Eq.~(\ref{mode2}) in the frame rotating at
$\Delta_2=w_m$ for the slow variables $\delta \tilde{o}(t) \equiv \delta o(t)\exp\{-iw_m t\}$ and neglect the
terms fastly oscillating at the frequency $2w_m$,
so to get \cite{Braunstein,pinard2}
\begin{equation}
\label{mode3} \delta \dot{\tilde{a}}_2=-\kappa_2\delta \tilde{a}_2 +i \frac{G_2 \alpha_2 }{\sqrt{2}} \delta \tilde{b} +\sqrt{2\kappa_2} \tilde{a}_{2}^{in}(t),
\end{equation}
where $\delta{b}=(i \delta p + \delta q)/\sqrt{2}$. If $\kappa_2 \gg G_2 |\alpha_2| /\sqrt{2}$, the cavity mode adiabatically follows the mirror
dynamics and one has
$
\delta \tilde{a}_2 \simeq i (G_2 \alpha_2/\kappa_2\sqrt{2}) \delta \tilde{b} +\sqrt{2/\kappa_2} \tilde{a}_{2}^{in}(t)
$. Using $\tilde{a}_2^{out}=\sqrt{2\kappa_2}\delta \tilde{a}_2-\tilde{a}_{2}^{in}$ \cite{milwal}, we finally get
\begin{equation}
\label{output} \tilde{a}_2^{out}= i \frac{G_2 \alpha_2 }{\sqrt{\kappa_2}} \delta \tilde{b} + \tilde{a}_{2}^{in}(t),
\end{equation}
showing that, in the chosen parameter regime, the output light of
$C_2$ gives a direct measurement of the mirror dynamics. By changing
the phases of the two local oscillators and by measuring the
correlations between the two cavity outputs one can determine all
the entries of the CM $V$ and from them numerically extract the
logarithmic negativity $E_{\mathcal{N}}$ by means of
Eq.~(\ref{logneg}).

In conclusion, we have shown that a Fabry-Perot cavity with an
oscillating micro-mirror and driven by coherent light can produce
robust and stationary entanglement between the optical intracavity
mode and the mechanical mode of the mirror. The amount of
entanglement is quantified by the logarithmic negativity and
surprisingly robust against increasing temperature: for experimental
parameters close to those of recently performed experiments
\cite{gigan06} entanglement may persist above 20K in the case of a
$5$ ng mechanical oscillator. Finally, we suggest a readout scheme
that allows a full experimental characterization of the CV Gaussian
steady state of the system and hence a measurement of the generated
entanglement.

We acknowledge helpful discussions with C. Brukner, J. Eisert, J.
Kofler, M. Paternostro. D. V. thanks the IQOQI in Vienna for
hospitality. This work was supported by the Austrian Science Fund
FWF (SFB15), by the European Commission under the programs RamboQ
and SECOQC and by the City of Vienna.

\end{document}